\documentclass[12pt]{article}
\usepackage{amsmath,amssymb,amsfonts,amsthm}
\usepackage[english]{babel}
\newcommand{\be}{\begin{equation}}
\newcommand{\ee}{\end{equation}}
\newcommand{\bea}{\begin{eqnarray}}
\newcommand{\eea}{\end{eqnarray}}
\newcommand{\nn}{\nonumber \\}
\newcommand{\p}[1]{(\ref{#1})}
\newcommand{\lb}{\label}

\usepackage{caption}
\usepackage{subcaption}

\begin{document}
\begin{titlepage}
\vspace*{1.5cm}
%\begin{flushright}
%JINR E2-2015-??
%\end{flushright}
%\vfill
%\vfill
\begin{center}
\baselineskip=16pt {\Large\bf Long multiplets in supersymmetric mechanics}

\vskip 0.3cm {\large {\sl }} \vskip 10.mm {\bf $\;$  E. Ivanov$^{\,a}$ and S. Sidorov$^{\,b}$
}
\vspace{1cm}

{\it Bogoliubov Laboratory of Theoretical Physics, JINR, \\
141980 Dubna, Moscow Region, Russia\\
}
\end{center}
\vfill

\par
\begin{center}
{\bf Abstract}
\end{center}

The ``long'' indecomposable ${\cal N}{=}2, d{=}1$ multiplet ${\bf (2, 4, 2)}$ defined in \texttt{arXiv:1503.\break 05537 [hep-th]}
as a deformation of the pair of chiral multiplets   ${\bf (2, 2, 0)}$ and ${\bf (0, 2, 2)}$ by a number of the mass-dimension parameters
is described in the superfield approach. We present its most general superfield and component actions, as well as a generalization
to the case with the superfields of the opposite Grassmann parities and dimensionless deformation parameter. We show that the long
${\cal N}{=}2, d{=}1$ multiplets are naturally embedded into the chiral $SU(2|1), d=1$ superfields having nonzero external spins
with respect to $SU(2) \subset SU(2|1)$. A superfield with spin $s$ contains $2s$ long multiplets and two short multiplets
${\bf (2, 2, 0)}$ and ${\bf (0, 2, 2)}$. Two possible ${\cal N}{=}4, d{=}1$ generalizations of the ${\cal N}=2$ long multiplet
in the superfield approach are also proposed.

\vspace{2.5cm}

\begin{quote}
\vfill \vfill \vfill \vfill \vfill \hrule width 5.cm \vskip 2.mm
{\small
\noindent $^a$ eivanov@theor.jinr.ru\\
\noindent $^b$ sidorovstepan88@gmail.com
}
\end{quote}
\end{titlepage}
\section{Introduction}
Supersymmetric quantum  mechanics (SQM) \cite{Witten} is the
simplest supersymmetric theory. One of the popular views of SQM is
that it provides the extreme version of the dimensional reduction of
the supersymmetric field theories and, as such, captures the salient
features of the latter, e.g., allows one to understand in depth the
conditions under which supersymmetry can be spontaneously broken.
On the other hand, $d=1$ supersymmetry and the associated models are
of obvious interest on their own, as the appropriate laboratory for
studying general peculiarities of supersymmetric theories and
constructing superextensions of black holes, the integrable
intrinsically one-dimensional systems like Calogero-Moser ones, etc.
(see \cite{superc} and references therein). Supersymmetry in
one dimension reveals particular features which distinguish it from its
higher-dimensional counterparts. For instance, some multiplets which
are on-shell in higher dimensions become off-shell after reducing
them to $d=1$ \cite{IKLech,DI}; there is such a specifically $d=1$
property as the automorphic duality relating multiplets
with the same number of fermionic fields but with different
divisions of the bosonic fields into the subsets of the physical and
auxiliary fields \cite{Autom,PT}, etc. Most of these peculiarities
naturally extend to various versions of the $SU(2|1)$ SQM as a
deformation of the standard SQM by an intrinsic mass parameter
\cite{weak,kahler}, \cite{DSQM} - \cite{DHSS}.

The aim of the present paper is to pay attention to the one more peculiar feature of $d=1$ supersymmetry, which manifests itself
in systems involving a few multiplets of similar sorts, and to give the superfield description of this new phenomenon. Namely, it turns out
that in a number of cases the sets of $d=1$ multiplets admit a deformation into new irreducible but indecomposable multiplets
called ``long multiplets'' in \cite{Casimir}. Actually, a similar possibility was noticed for the first time in \cite{Katona},
and we will present its superfield formulation too, (in Sec. 4). However, our main focus will be on the multiplets treated
in \cite{Casimir}. So, it is natural to start with a short account of what has been done there.

In that paper, $SU(2|1)$ supersymmetric mechanics was recovered by a dimensional reduction from the four-dimensional
Lagrangian of the chiral multiplet on the manifold $S^3\times\mathbb{R}$
\cite{FS}. After reduction, the $SU(2|1)$ supersymmetric system is
described by an infinite sum of $d=1$ supermultiplets. These multiplets
were treated in the framework of ${\cal N}=2$ supersymmetry with the
superalgebra \bea
    \lbrace Q, \bar{Q}\rbrace = 2(H - \Sigma)\,, \quad Q^2 = \bar Q^2 = 0\,, \label{algebraSigma}
\eea
where $H$ is the Hamiltonian and $\Sigma$ is some charge operator. It is assumed that both $H$ and $\Sigma$ (not only  $H -\Sigma$) commute
with the supercharges. The latter  requirement is related to the fact that the ${\cal N}=2$ superalgebra \p{algebraSigma} was defined in \cite{Casimir}
as a subalgebra of $su(2|1)$, with $H$ as the ``genuine'' Hamiltonian commuting with {\it all} $su(2|1)$ generators\footnote{This $H$ can be interpreted as
a {\it central charge} operator added to the standard $U(2)$ bosonic generators of $SU(2|1)$ \cite{FS,SKO}.} and $\Sigma$ as a combination of the remaining
bosonic $su(2|1)$ generators. The basic reason for the choice of $H$ as the Hamiltonian was the desire
to keep $SU(2|1)$ supercharges time independent. It allows one to Wick rotate the time coordinate and work in the {\it Lorentzian/Euclidean} signature. If we forget about this
$su(2|1)$ origin of \p{algebraSigma}, the presence of $\Sigma$ is not necessary since it can always be removed
by a field redefinition (see \p{sigma} below), and $H -\Sigma$ can be chosen as the Hamiltonian, with respect to which the spectrum is wittingly
bounded from below because of the evident condition $H -\Sigma \geq 0\,$.
However, as we are going to establish the relationships with the $SU(2|1)$ multiplets (Sec. 3), we explicitly keep $\Sigma$ in \p{algebraSigma}.

Proceeding from the superalgebra \p{algebraSigma}, the reference \cite{Casimir} revealed a new type of ${\cal N}=2$, $d=1$ multiplet
called a ``long'' multiplet. This multiplet involves some irreducible bosonic chiral ${\bf (2, 2, 0)}$ multiplet $(z, \xi)$, where
 $z(t)$ and $\xi(t)$ are bosonic and fermionic complex $d=1$ fields, as well as a fermionic ``quotient'' ${\bf (0, 2, 2)}$
 collecting the fermionic and bosonic
 fields $\pi(t)$ and $B(t)$. The ${\cal N}=2$ supersymmetry is realized on this set as
\bea
    &&\delta z  =-\,\sqrt{2}\,\epsilon\,\xi\,,\qquad\delta \xi =  \sqrt{2}\, i\,\bar{\epsilon}\,D_t z\,,\nn
    &&\delta \pi = -\,\sqrt{2}\,\epsilon B -\sqrt{2}\,\rho\,\bar{\epsilon}\,z\,,\qquad\delta B =
    \sqrt{2}\,i\,\bar{\epsilon}\,D_t{\pi}+\sqrt{2}\,\rho\,\bar\epsilon\,\xi\,,\label{long-tr}
\eea
where
\bea
    D_t = \partial_t + i\sigma\,, \qquad \bar{D}_t = \partial_t -  i\sigma \label{Dt}
\eea
and $\sigma$ and $\rho$ are some real parameters of the mass dimension. The factors $\sqrt{2}$ were introduced for further convenience.
The parameter $\rho$ is a deformation parameter responsible for combining the two chiral ``short'' ${\bf (2, 2, 0)}$
and ${\bf (0, 2, 2)}$ multiplets
into a ``long'' multiplet. In the limit $\rho=0$\,, the latter is split into the direct sum of chiral multiplets which transform under
${\cal N}=2, d=1$
supersymmetry independently of each other. Calculating the  Lie brackets of \eqref{long-tr}, one finds
that the generator $\Sigma$ is realized on each holomorphic component as $\Sigma = \sigma\,$, while $H=i\partial_t\,$. Redefining all fields as
\be
    z \rightarrow e^{-i\sigma t}z\,,\quad \xi \rightarrow e^{-i\sigma t}\xi\,,\quad
    \pi \rightarrow e^{-i\sigma t}\pi\,,\quad B \rightarrow e^{-i\sigma t}B\,,\qquad \mbox{c.c.},\label{sigma}
\ee
we can reduce the transformations \eqref{long-tr} to their $\sigma = 0$ case. This also leads to some shifting of the time-translation generator $H$, such that
on the new holomorphic fields it becomes $H = i\partial_t + \sigma$\,. Thus, the new time translation generator is identified with $H-\Sigma$\,; i.e.,
the redefinition \p{sigma} turns \eqref{algebraSigma} into the form of the standard ${\cal N}=2, d=1$ Poincar\'e superalgebra with the Hamiltonian $H-\Sigma = i\partial_t$\,.
So, at the level of ${\cal N}=2$ superalgebra, leaving aside the issue of its embedding into $su(2|1)$, the parameter $\sigma$ is unessential
and could be put equal to zero from the very beginning.
The parameter $\rho$ cannot be removed by any field redefinition, and it is the genuine deformation parameter. It is worth mentioning
that the long ${\cal N}=2$ multiplet has the same field content as a chiral ${\cal N}=4, d=1$
multiplet $({\bf 2, 4, 2})$.

The free Lagrangian, which is invariant, up to a total derivative, under \eqref{long-tr}, reads
\bea
    {\cal L}_{(\rho)}^{\rm free}&=& \bar{D}_t{\bar{z}}\,D_t{z}+ \frac{i}{2}\left(\bar{\xi}\,D_t\xi + \xi\,\bar{D}_t\bar{\xi}\,\right)
    +\frac{i}{2}\left(\bar{\pi}\,D_t\pi + \pi\,\bar{D}_t\bar{\pi}\right) +B\bar{B}\nn
    &&-\,\rho\left(\xi\bar{\pi}+\pi\bar{\xi}\,\right) - \rho^2 z\bar{z}-\gamma\left[\frac{i}{2}\left(z\,\bar{D}_t\bar{z}
    -\bar{z}\,D_t{z}\right) +\xi\bar{\xi}\right].\label{free}
\eea
The Lagrangian contains three independent parameters of mass dimension: $\sigma$\,,  $\rho$ and $\gamma$\,.
The external Wess-Zumino (WZ) term $\sim\gamma$ involves the irreducible set $(z\,, \xi)$ only. The whole effect of introducing $\rho \neq 0$
is the appearance of the new
oscillator-type terms for the physical bosonic and fermionic fields. The remaining parameters also contribute to
both these terms and the WZ term, as is seen after
rewriting \p{free} in a more detailed form:
\bea
    {\cal L}_{(\rho)}^{\rm free}&=& \dot{\bar z}\,\dot{z}+ \frac{i}{2}\,(\bar{\xi}\,\dot\xi + \xi\,\dot{\bar \xi}\,)
    +\frac{i}{2}\,(\bar{\pi}\,\dot\pi + \pi\,\dot{\bar \pi})- \rho\,(\xi\bar{\pi}+\pi\bar{\xi}\,) +
    (\sigma -\gamma)\,\xi\bar\xi + \sigma\, \pi\bar\pi\nn
    && + \,(\sigma^2 - \rho^2 -\gamma\sigma)z\bar{z} +\frac{i}{2}\,(\gamma - 2\sigma)\,(\bar z\,\dot{z}
    - {z}\,\dot{\bar z})  + B\bar{B}\,.\label{free1}
\eea

It is worth pointing out that the Lagrangian \p{free} consists of the three independent invariants. One of them 
is formed by the first two terms and is just
the free Lagrangian of the irreducible chiral $({\bf 2, 2, 0})$ multiplet $(z, \xi)$; the other one is the WZ term of this multiplet 
$\sim \gamma\,$; the remainder in \p{free} is
the invariant involving the $({\bf 0, 2, 2})$ quotient fields $(\pi, B)$. The relative coefficient between the first and third invariants 
has been fixed by requiring $\pi$
to have the standardly normalized kinetic term. In what follows, the various Lagrangians of the long multiplet will be chosen 
in such a way that their free parts
have the form \p{free}. In particular,  both the kinetic term $\dot {\bar z}\dot {z}$ and the auxiliary
field term $B\bar B$ will always enter with the same coefficient equal to 1.

The quantum brackets for the involved variables are
\bea
    \left[z,p_z\right]=i\,,\qquad\left[\bar{z},\bar{p}_{\bar{z}}\right]=i\,,\qquad
    \left\lbrace\xi ,\bar{\xi}\,\right\rbrace = 1\,,\qquad
    \left\lbrace\pi ,\bar{\pi}\right\rbrace = 1\,.
\eea
The corresponding quantum Weyl-ordered Hamiltonian reads
\bea
    H &=& p_z\bar{p}_{\bar{z}}-i\left(\sigma - \frac{\gamma}{2}\right)\left(zp_z-\bar{z}\bar{p}_{\bar{z}}\right)
    + \left(\rho^2+\frac{\gamma^2}{4}\right)z\bar{z}-\left(\sigma - \gamma\right)\,\xi\bar{\xi}-\sigma\,\pi\bar{\pi}\nn
    &&+\,\rho\left(\xi\bar{\pi}+\pi\bar{\xi}\,\right)-\frac{\gamma}{2}+\sigma.
    \label{H}
\eea
The quantum supercharges are
\bea
    Q=\sqrt{2}\,\xi\left(p_z -\frac{i}{2}\,\gamma\bar{z}\right)-\sqrt{2}\,i\rho\,\pi\,\bar{z}\,,\qquad
    \bar{Q}=\sqrt{2}\,\bar{\xi}\left(\bar{p}_{\bar{z}}+\frac{i}{2}\,\gamma z\right)+\sqrt{2}\,i\rho\,\bar{\pi}\,z\,.
\eea
The Weyl-ordered $U(1)$ generator reads
\bea
    \Sigma = -\,i\sigma\left(zp_z-\bar{z}\bar{p}_{\bar{z}}\right)-\sigma\left(\xi\bar{\xi}+\pi\bar{\pi}\right)+\sigma. \label{QuantSigma}
\eea

The quantum problem for the Hamiltonian \eqref{H} was solved in \cite{Casimir}.
The energy spectrum was shown to be bounded from below under the condition
\bea
    \sqrt{4\rho^2 + \gamma^2} \geq |\gamma - 2\sigma|\,,\lb{Bound}
\eea
and the energy of the ground state is zero, while all excited states possess the positive energy.\footnote{It was noted in \cite{Casimir}
that in the case of equality in \p{Bound}, i.e., $\sqrt{4\rho^2 + \gamma^2} = |\gamma - 2\sigma|\,,$
the ground state reveals an additional degeneracy.}  The reason why just \p{H} was taken in \cite{Casimir}
as the genuine Hamiltonian was explained earlier, after \p{algebraSigma}, and it is related to the embedding of
\p{algebraSigma} into the algebra $su(2|1)$. Note that we can pass in \eqref{algebraSigma} to the shifted generators
$H\rightarrow H + a$ and $\Sigma\rightarrow\Sigma + a$\,,
leaving the generator $H-\Sigma$ unaffected. Then both the energy spectrum with respect to $H$ and the value of the $\Sigma$ charge
are shifted by a constant $a$\,.
One can always choose $a$ in such a way that both bosonic generators are vanishing on the ground state.
Then, the condition \p{Bound} secures the absence of negative energies.

Staying solely within the ${\cal N}=2, d=1$ framework, we can always eliminate the parameter
$\sigma$ from the Lagrangian \p{free} through the redefinition \p{sigma} and pass to the difference of \p{H} and \p{QuantSigma} as the Hamiltonian,
in which all the dependence on $\sigma$ is canceled out and the condition \p{Bound} is identically satisfied (since we
can choose $\sigma =0$ altogether).

\setcounter{equation}{0}
\section{${\cal N}=2$ superfield description of long multiplets}
Here, we show how to reproduce the ${\cal N}=2$ supersymmetric mechanics of \cite{Casimir} from the ${\cal N}=2$ superfield approach
properly modified as compared to the standard one \cite{chiralN2}.

\subsection{Modified ${\cal N}=2, d=1$ superspace approach}
The ${\cal N}=2$, $d=1$ superspace coordinates $\left(t,\theta,\bar{\theta}\,\right)$ are defined to experience the standard  transformations
with respect to the superalgebra \p{algebraSigma}
\bea
    \delta\theta =\epsilon\,, \qquad \delta\bar{\theta}=\bar{\epsilon}\,,\qquad
    \delta t=i\left(\epsilon\,\bar\theta + \bar{\epsilon}\,\theta\right). \lb{CoordN2}
\eea
The ${\cal N}=2$ covariant derivatives ${\cal D}, \bar{\cal D}$ are defined by
\bea
    {\cal D}=\frac{\partial}{\partial\theta}-i\bar{\theta}\partial_t + \bar{\theta}\tilde{\Sigma}\,,\qquad
    \bar{\cal D}=-\frac{\partial}{\partial \bar{\theta}}+i\theta\partial_t - \theta \tilde{\Sigma}\,,\lb{DefDN2}
\eea
and satisfy the following anticommutation relations:
\bea
    \lbrace {\cal D}, \bar{{\cal D}}\rbrace = 2i\partial_t - 2\tilde{\Sigma}\,, \quad
    \lbrace {\cal D}, {\cal D}\rbrace = \lbrace \bar{\cal D}, \bar{\cal D}\rbrace =0\,.
\eea
The new conserved operator $\tilde{\Sigma}$ does not act on the superspace coordinates and commutes with  the spinor derivatives
but can take nonzero values on the corresponding superfields. So it can be viewed as an active central charge \cite{Gates}.
The Hamiltonian is realized  as the time translations generator, $H=i\partial_t\,$.

The covariant derivatives possess nonstandard transformation properties under \p{CoordN2}
\bea
\delta  {\cal D} = \bar\epsilon\,\tilde{\Sigma}\,, \qquad \delta  \bar{\cal D} = -\,\epsilon\,\tilde{\Sigma}\,. \lb{TranN2D}
\eea
As was already mentioned, the general ${\cal N}=2$ superfield can have an external charge with respect to the generator $\tilde{\Sigma}$ and undergo the
 following (passive) transformation law
 \be
 \delta \Phi = \left(\epsilon\,\bar{\theta}+\bar{\epsilon}\,\theta\right)\tilde{\Sigma}\Phi\,. \lb{ModTransf1}
 \ee
The covariant spinor derivatives of such a superfield are again superfields, just owing to the transformation law \p{TranN2D}.
The spinor generators corresponding to the transformation law \p{ModTransf1} are defined as
\be
Q = -\frac{\partial}{\partial\theta}-i\bar{\theta}\partial_t + \bar{\theta}\tilde{\Sigma}\,, \quad \bar Q =
 -\frac{\partial}{\partial \bar{\theta}} - i\theta\partial_t + \theta \tilde{\Sigma}\,, \quad \{Q, \bar Q\}
 = 2i\partial_t - 2\tilde{\Sigma}\,.
\ee
As expected, they anticommute with the covariant spinor derivatives \p{DefDN2}.\footnote{By the similarity transformation
$\Phi = e^{-it\tilde{\Sigma}}\hat{\Phi}$, one could transform the covariant derivatives and supercharges to the standard form,
with $\tilde{\Sigma}$ becoming some extra $U(1)$ generator times a mass-dimension parameter.
In components,  it amounts just to the redefinition \eqref{sigma}.
However, the time translation generator in the realization on $\hat{\Phi}$
necessarily involves the $U(1)$ rotation part $\sim \tilde{\Sigma}$\,. We will  work in the original frame, where the time translation
generator is a pure shift, $H=i\partial_t$\,.}
\subsection{Long multiplet}
The long multiplet is described by the pair of fermionic and bosonic ${\cal N}=2$ superfields $\Psi$ and $Z$
which possess  nonzero $U(1)$ charge,
\be
\tilde{\Sigma}\Psi = \sigma\Psi\,, \quad \tilde{\Sigma}Z = \sigma Z\,,\lb{Eigen}
\ee
and are subject to the following constraints with $\rho \neq 0$:
\bea
    \bar{{\cal D}} \Psi = -\,\sqrt{2}\,\rho\,Z\quad\Rightarrow\quad \bar{{\cal D}} Z = 0\,.\label{NewLp}
\eea
The superfield $Z$ is chiral and describes an irreducible multiplet $({\bf 2,2,0})$, while $\Psi$, as a solution  of \p{NewLp},
can be represented as
\be
    \Psi = \Pi + \sqrt{2}\,\rho\,\bar\theta Z\,,\qquad \bar{{\cal D}}\Pi=0\,, \quad \tilde{\Sigma}\,\Pi = \sigma\Pi\,.
\ee
The component expansion of the chiral superfields $\Pi$ and $Z$ is defined by
\bea
    &&\bar{{\cal D}}Z=0\,,\qquad \tilde{\Sigma}\,Z = \sigma Z \qquad\Rightarrow\qquad Z=z + \sqrt{2}\,\theta\,\xi
    - i\,\theta\bar{\theta}\,D_t z\,,\nn
    &&\bar{{\cal D}}\Pi=0\,,\qquad \tilde{\Sigma}\,\Pi = \sigma\Pi \qquad\Rightarrow\qquad \Pi = \pi +
    \sqrt{2}\,\theta B - i\,\theta\bar{\theta}\,D_t\pi\,.\label{chiral}
\eea
Under ${\cal N}=2$ supersymmetry, the superfields $\Psi$ and $Z$ transform with a nontrivial phase factor $\propto \sigma$\,:
\bea
    \delta\Psi = \sigma \left(\epsilon\,\bar{\theta}+\bar{\epsilon}\,\theta\right)\Psi\,, \quad
    \delta Z= \sigma \left(\epsilon\,\bar{\theta}+\bar{\epsilon}\,\theta\right)Z\,.\label{trSF1}
\eea
These transformations imply the following transformation law for the superfield $\Pi$:
\bea
    \delta\Pi = \sigma \left(\epsilon\,\bar{\theta}+\bar{\epsilon}\,\theta\right)\Pi - \sqrt{2}\,\rho\,\bar{\epsilon}\,Z\,.
\eea
For the component fields in \p{chiral}, these superfield transformation laws induce just the transformations \eqref{long-tr}.

Thus, we started with the general complex bosonic and fermionic ${\cal N}=2$ superfields $Z$ and $\Psi$ having the transformation
properties \p{trSF1} (we can equally choose $\sigma \neq 0$ or $\sigma =0$; this is not crucial for our consideration).
Then, we impose on $Z$ the standard chirality condition, which makes it the chiral ${\cal N}=2$ superfield carrying
the irreducible multiplet $({\bf 2, 2, 0})$.
The  pivotal point is that the fermionic superfield $\Psi$ is now subjected, instead of the standard chirality condition,
to the new condition given in \p{NewLp}. It expresses some components of  $\Psi$ in terms of the components of $Z$
and forces the remaining superfield
$\Pi$ with the field content $({\bf 0, 2, 2})$ to transform through $Z$. The nontrivial pieces in the transformations
\eqref{long-tr} of the components of $\Pi$ arise
just as a consequence of this new manifestly ${\cal N}=2$ supersymmetric constraint on $\Psi$.
In the limit $\rho=0$\,, the superfields \eqref{chiral} describe two independent irreducible ${\cal N}=2$ multiplets,
$({\bf 2, 2, 0})$ and $({\bf 0, 2, 2})$.
In other words, in this limit, the indecomposable long ${\cal N}=2$ multiplet $({\bf 2, 4, 2})$  splits into the direct sum of two ``short''
irreducible  ${\cal N}=2$ multiplets. At $\rho \neq 0$, such a splitting cannot be accomplished by any field
redefinition.\footnote{It is worth mentioning that the reduction $Z=const, \sigma =0$ in \p{NewLp}
yields a modified version of the multiplet $({\bf 0,2,2})$ considered in \cite{LeTo}.}

The most general Lagrangian of the long multiplet is a sum of two invariant Lagrangians,
\bea
    {\cal L}_{(\rho)} = {\cal L}_{(Z)} + {\cal L}_{(\Psi, Z)}\,,\label{Lp}
\eea
where the invariant Lagrangian ${\cal L}_{(\Psi, Z)}$ is
\bea
    {\cal L}_{(\Psi,Z)} = \frac{1}{2}\int d\bar{\theta}\, d\theta\,\Psi\bar{\Psi}\,h_1\left(Z,\bar{Z}\right)
\eea
and ${\cal L}_{(Z)}$ is the general Lagrangian for the chiral superfield $Z$ \cite{chiralN2}:
\bea
    {\cal L}_{(Z)} = \frac{1}{4}\int d\bar{\theta}\, d\theta\,\bar{\cal D}\bar{Z}\,{\cal D}Z\,h_0\left(Z,\bar{Z}\right)
    -\frac{\gamma}{2}\int d\bar{\theta}\, d\theta\,h_{(\gamma)}\left(Z,\bar{Z}\right).
\eea
The component form of the Lagrangian ${\cal L}_{(\rho)}$ reads
\bea
    {\cal L}_{(\rho)} &=& h_0\,D_t{z}\,\bar{D}_t{\bar{z}}+\frac{i}{2}\,h_0\left(\bar{\xi}\,D_t\xi - \bar{D}_t\bar{\xi}\,\xi\right)
    +\frac{i}{2}\,\xi\bar{\xi}\left(\bar{D}_t{\bar{z}}\,\partial_{\bar{z}}h_0-D_t{z}\,\partial_{z}h_0\right) \nn
    && -\,\gamma\left[\frac{i}{2}\left(\bar{D}_t\bar{z}\,\partial_{\bar{z}}h_{(\gamma)} - D_t{z}\,\partial_{z}h_{(\gamma)}\right)
    + \partial_{z}\partial_{\bar{z}}h_{(\gamma)}\,\xi\bar{\xi}\right]+\frac{i}{2}\,h_{1}\left(\bar{\pi}\,D_t\pi - \bar{D}_t\bar{\pi}\,\pi \right) \nn
    && +\, B\bar{B}\,h_{1}
    +\xi\pi\,\bar{B}\,\partial_{z}h_{1} + \bar{\pi}\bar{\xi}\,B\, \partial_{\bar{z}}h_{1}+\frac{i}{2}\,
    \pi\bar{\pi}\left(\bar{D}_t{\bar{z}}\,\partial_{\bar{z}}h_{1}-D_t{z}\,\partial_{z}h_{1}\right)\nn
    &&+\,\pi\bar{\pi} \,\xi\bar{\xi}\,\partial_{z}\partial_{\bar{z}}h_{1}-\rho
    \left(\pi\bar{\xi}\,\bar{z}\partial_{\bar{z}}h_{1}+\xi\bar{\pi}\,z \partial_{z}h_{1}\right)
    -\rho\left(\pi\bar{\xi}+\xi\bar{\pi}\right)h_{1} - \rho^2 z\bar{z}\, h_{1}.\lb{LagrLongComp}
\eea
In the case of $\sigma \neq 0$\,, supersymmetry in addition requires the $U(1)$ invariance of the above Lagrangians.
According to \eqref{trSF1}, the arbitrary functions $h_0\left(z,\bar{z}\right)$, $h_{(\gamma)}\left(z,\bar{z}\right)$, and $h_1\left(z,\bar{z}\right)$
must satisfy the $U(1)$ invariance conditions in this case:
\bea
    h_0\left(z,\bar{z}\right)\equiv h_0\left(z\bar{z}\right),\;\;h_{(\gamma)}\left(z,\bar{z}\right)\equiv h_{(\gamma)}\left(z\bar{z}\right),\;\;h_1
    \left(z,\bar{z}\right)\equiv h_1\left(z\bar{z}\right),\quad\sigma \neq 0\,.\label{U1}
\eea
However, making the redefinition \p{sigma} in the $U(1)$ invariant Lagrangian, we can
eliminate  any dependence on $\sigma$ in it, after which it becomes a particular $U(1)$ invariant case
of \p{LagrLongComp} with $\sigma =0$. Thus, the most general invariant Lagrangian of the long multiplet is just the $\sigma =0$
version of \p{LagrLongComp}. No additional $U(1)$ invariance is required.

We see that the deformation parameter $\rho$ is responsible for the appearance of the new potential and Yukawa-type terms.

As an example, we present the superfield form of the free Lagrangian \eqref{free}:
\bea
    {\cal L}_{(\rho)}^{\rm free}&=&\frac{1}{4}\int d\bar{\theta}\, d\theta\left(\bar{\cal D}\bar{Z}\,{\cal D}Z + 2\Psi\bar{\Psi}
    - 2 \gamma\,Z\bar{Z}\right).
\eea
It is a sum of three superfield invariants.

Finally, we note that the constraint for $\Psi$ in \p{NewLp} with $\rho \neq 0$ looks like the expression
of the chiral superfield $Z$ through
an unconstrained  fermionic prepotential $\Psi$. However, in the present case, this analogy does not work because the prepotential
should admit the appropriate
pregauge freedom like $\delta \Psi = \bar{\cal D}\Lambda$, while the  Lagrangians in \p{Lp} do not respect such a freedom. As a result,
the extra fields $\pi(t), B(t)$ accommodated by the $({\bf 0, 2, 2})$ superfield $\Pi$ in \p{chiral} are by no means gauge degrees of freedom.

\subsection{Long multiplet with the twisted Grassmann parity}
The superfields $\Psi$ and $Z$ subjected to the constraints \eqref{NewLp} have their opposite Grassmann parity counterparts
$Y$ (bosonic) and $\Omega$ (fermionic) satisfying the constraints
\bea
    \bar{{\cal D}} Y = -\,\sqrt{2}\,\varrho\,\Omega\quad\Rightarrow\quad \bar{{\cal D}} \Omega = 0\,, \qquad \tilde{\Sigma}\,Y=\sigma Y\,, \;
    \tilde{\Sigma}\,\Omega=\sigma \Omega\,.
\eea
The parameter $\varrho$, in contrast to $\rho$ in \eqref{NewLp}, is dimensionless. Obviously, $Y$ can be written as
\bea
    Y = Y_0 + \sqrt{2}\,\varrho\,\bar\theta\,\Omega\,, \qquad \bar{{\cal D}}Y_0=0\,,\quad \tilde{\Sigma}\,Y_0 = \sigma Y_0\,,
\eea
with
\bea
Y_0=y + \sqrt{2}\,\theta\,\eta - i\,\theta\bar{\theta}\,D_t y\,,\qquad
    \Omega = \chi + \sqrt{2}\,\theta\,C - i\,\theta\bar{\theta}\,D_t\chi\,.\lb{CompY}
\eea

The ${\cal N}=2$ transformations of $Y$ and $\Omega$ are given by
\bea
    \delta Y = \sigma \left(\epsilon\,\bar{\theta}+\bar{\epsilon}\,\theta\right)Y\,,
    \qquad
    \delta \Omega = \sigma \left(\epsilon\,\bar{\theta}+\bar{\epsilon}\,\theta\right)\Omega\,.
\eea
They induce the following transformations for the component fields defined in \p{CompY}:
\bea
    &&\delta y  =-\,\sqrt{2}\,\epsilon\,\eta - \sqrt{2}\,\varrho\,\bar{\epsilon}\,\chi\,,\qquad
    \delta \eta =  \sqrt{2}\, i\,\bar{\epsilon}\,D_t y +\sqrt{2}\,\varrho\,\bar{\epsilon}\,C\,,\nn
    &&\delta \chi = -\,\sqrt{2}\,\epsilon\,C\,,\qquad\delta C = \sqrt{2}\,i\,\bar{\epsilon}\,D_t{\chi}\,.
\eea
\subsubsection{Free model}
For simplicity, we consider the free action with $\sigma = 0$\,:
\bea
    S_{(\varrho)}^{\rm free} = \frac{1}{4}\int dt\, d\bar{\theta}\, d\theta\left(\bar{\cal D}\bar{Y}\,{\cal D}Y + 2\,\Omega\bar{\Omega}
    - 2\gamma\,Y\bar{Y}\right).
\eea
In the component notation, it reads
\bea
    {\cal L}_{(\varrho)}^{\rm free}&=& \dot{\bar{y}}\dot{y}+ \frac{i}{2}\left(\bar{\chi}\dot\chi + \chi\dot{\bar{\chi}}\right)
    +\frac{i}{2}\left(\bar{\eta}\dot\eta + \eta\dot{\bar{\eta}}\right)
    +i\varrho\left(\bar{C}\dot{y}-C\dot{\bar{y}}\right) +\left(1+\varrho^2\right)C\bar{C}\nn
    &&-\,\gamma\left[\frac{i}{2}\left(y\,\dot{\bar{y}} -\bar{y}\dot{y}\right) +\eta\bar{\eta}-\varrho^2\chi\bar{\chi}
    -\varrho\left(C\bar{y}+\bar{C}y\right)\right].
\eea
We eliminate the  auxiliary fields  by their equations of motion,
\bea
    C = \frac{\varrho}{1+\varrho^2}\left(-\,i\dot{y}-\gamma y\right),\qquad
    \bar{C}= \frac{\varrho}{1+\varrho^2}\left(i\dot{\bar{y}}-\gamma\bar{y}\right),
\eea
and obtain the on-shell Lagrangian in the form\footnote{The constant metric factor in the first line can be removed by rescaling $y, \bar y$.}
\bea%\left.\right|_{\rm on-shell}
    {\cal L}_{(\varrho)}^{\rm free}&=& \frac{1}{1+\varrho^2}\left[\dot{\bar{y}}\dot{y}-\frac{i}{2}\,\gamma\left(1-\varrho^2\right)
    \left(y\,\dot{\bar{y}} -\bar{y}\dot{y}\right) -\gamma^2\varrho^2 y\bar{y}\right]\nn
    &&+\,\frac{i}{2}\left(\bar{\chi}\dot\chi + \chi\dot{\bar{\chi}}\right)
    +\frac{i}{2}\left(\bar{\eta}\dot\eta + \eta\dot{\bar{\eta}}\right)-\gamma\,\eta\bar{\eta}+\gamma\varrho^2\chi\bar{\chi}\,.\label{onshell}
\eea
The relevant on-shell transformations are
\bea
    &&\delta y  =-\,\sqrt{2}\,\epsilon\,\eta - \sqrt{2}\,\varrho\,\bar{\epsilon}\,\chi\,,\nn
    &&\delta \eta =  \frac{\sqrt{2}\,\bar{\epsilon}}{1+\varrho^2}\left(i\dot{y}-\gamma \varrho^2 y\right),\qquad
    \delta \chi = \frac{\sqrt{2}\,\varrho\,\epsilon}{1+\varrho^2}\left(i\dot{y}+\gamma y\right).
\eea

Quantizing the free model \eqref{onshell} as
\bea
    \left[y,p_y\right]=i\,,\qquad\left[\bar{y},\bar{p}_{\bar{y}}\right]=i\,,\qquad
    \left\lbrace\eta ,\bar{\eta}\,\right\rbrace = 1\,,\qquad
    \left\lbrace\chi ,\bar{\chi}\right\rbrace = 1\,,
\eea
we write the quantum Hamiltonian,
\bea
    H &=& \frac{1}{1+\varrho^2}\left[\left(1+\varrho^2\right)\bar{p}_{\bar{y}}+\frac{i}{2}\,\gamma\left(1-\varrho^2\right)y\right]
    \left[\left(1+\varrho^2\right)p_y-\frac{i}{2}\,\gamma\left(1-\varrho^2\right)\bar{y}\right] \nn
    &&+\,\frac{\gamma^2\varrho^2}{1+\varrho^2}\,y\bar{y}+\gamma\,\eta\bar{\eta}-\gamma\varrho^2\chi\bar{\chi}\,,
\eea
and the quantum supercharges,
\bea
    &&Q=\sqrt{2}\left[\eta\left(p_y -\frac{i}{2}\,\gamma\bar{y}\right)+\varrho \bar{\chi}\left(\bar{p}_{\bar{y}}-\frac{i}{2}\,\gamma y\right)\right],\nn
    &&\bar{Q}=\sqrt{2}\left[\bar{\eta}\left(\bar{p}_{\bar{y}}+\frac{i}{2}\,\gamma y\right)+ \varrho \chi\left(p_y +\frac{i}{2}\,\gamma\bar{y}\right)\right].
\eea
%Since $\sigma=0$\,, the relevant $U(1)$ generator is vanishing: $\Sigma = 0$\,.

We observe that the nonzero deformation parameter $\varrho$ gives rise to the appearance of the new oscillator-type terms
and renormalizes the strength of the external ``magnetic field'' in the WZ term. Note that these new terms are proportional
to $\gamma$ and so appear only on account of adding the external WZ term.  In the general case, the new terms
also appear only due to some superfield interactions.

\setcounter{equation}{0}
\section{$SU(2|1)$ chiral multiplets}
In this section, we discuss how long ${\cal N}=2, d=1$ multiplets can be embedded into the $SU(2|1)$ chiral multiplets in the framework
of the $SU(2|1)$ mechanics as a deformation of the ${\cal N}=4$ mechanics \cite{DSQM}. We will proceed from  the manifestly $SU(2|1)$
covariant world-line superfield approach.

The relevant deformed superalgebra is the centrally extended superalgebra $\widehat{su}(2|1)$:
\bea
    &&\lbrace Q^{i}, \bar{Q}_{j}\rbrace = 2m I^i_j + 2\delta^i_j\left(H-mF\right) ,\qquad\left[I^i_j,  I^k_l\right]
    = \delta^k_j I^i_l - \delta^i_l I^k_j\,,\nn
    &&\left[I^i_j, \bar{Q}_{l}\right] = \frac{1}{2}\,\delta^i_j\bar{Q}_{l}-\delta^i_l\bar{Q}_{j}\, ,\qquad \left[I^i_j, Q^{k}\right]
    = \delta^k_j Q^{i} - \frac{1}{2}\,\delta^i_j Q^{k},\nn
    &&\left[F,\bar{Q}_{l}\right]=-\frac{1}{2}\,\bar{Q}_{l}\,,\qquad \left[F, Q^{k}\right] = \frac{1}{2}\,Q^{k}.\label{algebra}
\eea
The supersymmetric $\epsilon$ transformations of the superspace coordinates $\zeta := \{t,\theta_i,\bar{\theta}^k\}$,
$\bar{\theta}^i=\overline{\left(\theta_i\right)}$ are given by
\bea
    &&\delta\theta_{i}=\epsilon_{i} +
    2m\,\bar{\epsilon}^k\theta_k\theta_{i}\,,\quad
    \delta \bar{\theta}^{i}=\bar{\epsilon}^{i} -
    2m\,\epsilon_k\bar{\theta}^k\bar{\theta}^{i}\,,\quad
    \delta t = i\left(\bar{\epsilon}^k\theta_k + \epsilon_k\bar{\theta}^k\right).\lb{TransfCoord}
\eea

The $SU(2|1)$ covariant derivatives are defined as\footnote{We use the following conventions:
$\left(\theta\right)^2 =  \theta_i\theta^i$ and $\left(\bar\theta\,\right)^2 = \bar{\theta}^i\bar{\theta}_i$\,.}
\bea
    {\cal D}^i &=& \left[1+{m}\,\bar{\theta}^k\theta_k
    -\frac{3m^2}{8} \left(\theta\right)^2\left(\bar{\theta}\,\right)^2\right]\frac{\partial}{\partial\theta_i}
    - {m}\,\bar{\theta}^i\theta_j\frac{\partial}{\partial\theta_j}-i\bar{\theta}^i \partial_t\nn
    &&+\,m\,\bar{\theta}^i \tilde{F}- {m}\,\bar{\theta}^j\left(1 -m\,\bar{\theta}^k\theta_k \right)\tilde{I}^i_j\,,\nn
    \bar{{\cal D}}_j &=& -\left[1+ {m}\,\bar{\theta}^k\theta_k
    -\frac{3m^2}{8} \left(\theta\right)^2\left(\bar{\theta}\,\right)^2\right]\frac{\partial}{\partial\bar{\theta}^j}
    + {m}\,\bar{\theta}^k\theta_j\frac{\partial}{\partial\bar{\theta}^k}+i\theta_j\partial_t\nn
    &&-\,m\,\theta_j\tilde{F}+ {m}\,\theta_l\left(1 -m\,\bar{\theta}^k\theta_k \right)\tilde{I}^l_j\,,\label{cov}
\eea
where $\tilde{F}$ and $\tilde{I}^i_k$ are the ``matrix'' parts of the generators $F$ and $I^i_k$\,.
The $SU(2|1)$ superfields can carry an external $SU(2)$ spin index and $U(1)$ charge corresponding to these matrix parts.

The $SU(2|1)$ superfields can be subject to the chiral or antichiral conditions
\bea
   \bar{\cal D}_i \Phi =0\,,\qquad {\cal D}^i\bar{\Phi}=0\,.
\eea
They can be solved through the shortened superfields  living on the chiral subspaces
$\zeta_L=\left\lbrace t_L,\theta_i\right\rbrace$ and $\zeta_R=\left\lbrace t_R, \bar{\theta}^i\right\rbrace\,$, with
\bea
    t_L = t +i\,\bar{\theta}^k\theta_k -\frac{i}{2}\,m\left(\theta\right)^2\left(\bar{\theta}\,\right)^2, \qquad {\rm and \;\;c.c.}. \label{left}
\eea
These subspaces are closed under the transformations \p{TransfCoord}. The covariant derivatives $\bar{{\cal D}}_j$
can be rewritten in the basis $\left\lbrace t_L,\theta_i,\bar{\theta}_k\right\rbrace$ as
\bea
    \bar{{\cal D}}_j &=& -\left[1+ {m}\,\bar{\theta}^k\theta_k
    -\frac{3m^2}{8} \left(\theta\right)^2\left(\bar{\theta}\,\right)^2\right]\frac{\partial}{\partial\bar{\theta}^j}
    + {m}\,\bar{\theta}^k\theta_j\frac{\partial}{\partial\bar{\theta}^k}\nn
    &&-\,m\,\theta_j\tilde{F}+ {m}\,\theta_l\left(1 -m\,\bar{\theta}^k\theta_k \right)\tilde{I}^l_j\,.\label{covL}
\eea

The most general $SU(2|1)$ chiral multiplet defined in \cite{DSQM} is described by a chiral superfield $\Phi_A$ having an external index $A$ of $SU(2)\subset SU(2|1)$
and carrying an external $U(1)$ charge, $\tilde{F}\Phi_A = 2\kappa \Phi_A$. So,
it is characterized by the pair of real numbers $(s, 2\kappa)$\,, where $s$ is the external $SU(2)$ spin and $2\kappa$ is the external $U(1)$ charge.
As compared to the $SU(2)$ singlet chiral superfields ($s=0$) basically considered in \cite{DSQM}, the number of component fields in
$\Phi_A$ carrying the nonzero external spins $s=1/2, 1, \ldots $ increases  according to $${\bf\left( 2[2s+1], 4[2s+1], 2[2s+1]\right)}.$$

As was shown in \cite{DSQM}, the decomposition of the $s=0$ chiral superfield into the ${\cal N}=2$ multiplets
is given by a direct sum of chiral multiplets $({\bf 2, 2, 0})$ and $({\bf 0, 2, 2})$.\footnote{In \cite{DSQM},
the $SU(2|1)$ Lagrangians were reformulated in terms of ${\cal N}=2$ superfields pertinent to the standard definition of
the ${\cal N}=2$ spinor derivatives (without terms $\sim \tilde{\Sigma}$ in \p{DefDN2}).
Here, we deal with the centrally extended superalgebra \eqref{algebraSigma} and superfields having some external charges with respect
to the central charge generator $\Sigma$\,.} It turns out that the analogous decompositions of the $s\neq 0$ chiral superfields necessarily
involve the long
${\cal N}=2$ multiplets with $\rho \sim m$\,. Now, we will explicitly demonstrate this on the example of the $s=1/2$ chiral superfield.

\subsection{$s=1/2$}
%Speculating on this, let us consider the case $s=1/2$\,.
The $SU(2|1)$ chiral superfield $\Phi_i$ ($i=1,2$) in the $U(2)$ representation $\left(1/2 , 2\kappa\right)$ is defined by the constraints
\bea
    \bar{{\cal D}}_j\Phi_i = 0\,,\qquad \tilde{I}^{k}_{l}\Phi_i = \delta^k_i\Phi_l
    - \frac{1}{2}\,\delta^k_l\Phi_i\,,\qquad \tilde{F}\Phi_i = 2\kappa\,\Phi_i\,.
\eea
Using the explicit form of the covariant derivative \eqref{covL}, the chirality condition is written as
\bea
   && \left\lbrace\left[1+ {m}\,\bar{\theta}^k\theta_k
    -\frac{3m^2}{8} \left(\theta\right)^2\left(\bar{\theta}\,\right)^2\right]\frac{\partial}{\partial\bar{\theta}^j}
    - {m}\,\bar{\theta}^k\theta_j\frac{\partial}{\partial\bar{\theta}^k}\right\rbrace\Phi_i +\,2\kappa m\,\theta_j \Phi_i \nonumber\\
   &&  - m\,\theta_l\left(1 -m\,\bar{\theta}^k\theta_k \right)\left(\delta^l_i\Phi_j
    - \frac{1}{2}\,\delta^l_j\Phi_i\right)= 0\,,
\eea
and it is solved by
\bea
    \Phi_i\left(t_L,\theta_i,\bar{\theta}_k\right) &=& \left(1 + 2m\,\bar{\theta}^l\theta_l\right)^{-\kappa}
    \left[1 - \frac{3m^2}{16}\left(\bar{\theta}\,\right)^2\left(\theta\right)^2\right]\phi_{i}\left(t_L,\theta_i\right) \nn
    &&-\,m\left(\frac{1}{2}\,\delta^j_i\,\bar{\theta}^k\theta_k -\bar{\theta}^j\theta_i\right)\phi_j\left(t_L,\theta_i\right),\nn
    \phi_{i}\left(t_L,\theta_i\right) &=& z_i + \theta_i\psi - \sqrt{2}\,\theta^k\psi_{(ik)}+ \theta_k\theta^k B_i\,,
\eea
where
\bea
    \overline{\left(z_i\right)}=\bar{z}^i,\qquad\overline{\left(B^i\right)}=\bar{B}_i,\qquad\overline{\left(\psi\right)}
    =\bar{\psi},\qquad\overline{\left(\psi^{(ik)}\right)}=\bar{\psi}_{(ik)}\,.
\eea
The superfields $\Phi_i$ and $\phi_i$ transform as
\bea
    \delta\Phi_i &=& m\left(1 - m\,\bar{\theta}^l\theta_l\right)\left[\frac{1}{2}\,\delta^j_i
    \left(\epsilon_k\bar{\theta}^k+\bar{\epsilon}^k\theta_k\right)-\left(\epsilon_i\bar{\theta}^j+\bar{\epsilon}^j\theta_i\right)\right]\Phi_j\nn
    &&+\, 2\kappa m \left(\epsilon_k\bar{\theta}^k +\bar{\epsilon}^k\theta_k\right) \Phi_i\,,\nn
    \delta\phi_i &=& 4\kappa m\,(\bar{\epsilon}^k\theta_k)\,\phi_i + 2m\left(\frac{1}{2}\,\delta^j_i\,\bar{\epsilon}^k\theta_k
    - \bar{\epsilon}^j\theta_i\right)\phi_j\,.\label{trPhi}
\eea
The relevant $SU(2|1)$ transformations of the component fields are
\bea
    &&\delta z^i  =-\,\epsilon^i\psi -\sqrt{2}\,\epsilon_k\psi^{(ik)},\nn
    &&\delta \psi =  \bar{\epsilon}^k\left(i\,\nabla_t z_k + \frac{3m}{2}\,z_k\right)-\epsilon^k B_k\,,\nn
    &&\delta \psi^{(ik)} = \sqrt{2}\, \bar{\epsilon}^{(k}\left[i\,\nabla_t z^{i)} - \frac{m}{2}\,z^{i)}\right]-\sqrt{2}\,\epsilon^{(i} B^{k)},\nn
    &&\delta B^i = -\,\bar{\epsilon}^i\left(i\,\nabla_t\psi - \frac{m}{2}\,\psi\right)-\sqrt{2}\,\bar{\epsilon}_k
    \left(i\,\nabla_t\psi^{(ik)}+\frac{3m}{2}\,\psi^{(ik)}\right),
    \label{new-su21}
\eea
where
\bea
    \nabla_t = \partial_t +2 i\kappa m\,, \qquad \bar{\nabla}_t = \partial_t - 2 i\kappa m\,.\label{nabla}
\eea

Singling out, in \p{new-su21}, the subset of ${\cal N}=2$ transformations associated with the parameter $\epsilon_1\equiv\epsilon$,
we can identify the component fields $\left(z_i,\psi^{(ik)},\psi, B^i\right)$ with the system of three ${\cal N}=2$ multiplets:
one long multiplet and the irreducible chiral multiplets $({\bf 2, 2, 0})$ and $({\bf 0, 2, 2})$\,.
The long multiplet is formed by the fields $\left(z_1,\psi^{(12)},\psi, B^1\right)$:
\bea
z=z_1\,,\qquad\xi=\frac{\psi}{\sqrt{2}}-\psi_{(12)}\,,\qquad\pi=\frac{\psi}{\sqrt{2}}+\psi_{(12)}\,,\qquad B=-B^1.\label{subsets}
\eea
The parameter $\rho$ is equal to $-m$, and the charge $\sigma$ is identified with $\left(2\kappa-1/2\right)m$\,. Under these identifications, the $\epsilon_1, \bar\epsilon^1$
transformations from \p{new-su21} for the fields defined in \p{subsets} precisely coincide with \p{long-tr}.
The ${\cal N}=2$-irreducible multiplets $({\bf 2, 2, 0})$ and $({\bf 0, 2, 2})$ are composed of the fields $(z_2, \psi^{(11)})$ and $( \psi^{(22)},B^2)$.

To clarify this decomposition into ${\cal N}=2$ multiplets, let us note that the generator $\Sigma$ in the superalgebra \eqref{algebraSigma}
treated as a subalgebra of \eqref{algebra} is identified with the  combination of the $SU(2|1)$ bosonic generators
\bea
    \Sigma = m\left(F-I^1_1\right),\label{Sigma}
\eea
corresponding to the identification $Q\equiv Q^1$, $\bar{Q}\equiv\bar{Q}_1$ in \eqref{algebra}.
We see that the superalgebra \eqref{algebraSigma} proves properly embedded into \eqref{algebra}, with the same Hamiltonian $H$
which commutes with all $SU(2|1)$ supercharges.
%So, the following redefinition of the Hamiltonian as $H \rightarrow H - \Sigma$ reduces
%the $SU(2|1)$ supersymmetry to the standard ${\cal N}=2$ Poincar\'e supersymmetry.}
The generators $F$ and $I^1_1$ have the following realizations on the component fields:
\begin{center}
$
\begin{array}{|c|c|c|c|c|c|c|c|c|}
\hline
                & z_1 & z_2     & \psi & \psi^{(12)} & \psi^{(11)} & \psi^{(22)} & B^1& B^2\\
\hline
    I^1_1       & 1/2 & -1/2    & 0 & 0 & -1 & 1 & -1/2 & 1/2\\
\hline
    F - 2\kappa & 0 & 0     & -1/2 & -1/2 & -1/2 & -1/2 & -1 & -1\\
\hline
\end{array}$
\end{center}
The table shows also that the $U(1)$ charge $\Sigma$ takes the values $\left(2\kappa-1/2\right)m$\,, $\left(2\kappa-3/2\right)m$ and $\left(2\kappa+1/2\right)m$
on different irreducible ${\cal N}=2$ multiplets forming the chiral $SU(2|1)$ multiplet we are considering.
From the table, one also finds the $U(2)$ assignment of the component fields of the chiral multiplet $\left(s=1/2 , 2\kappa\right)$:
\bea
    {\rm bosons}\qquad &&\left(1/2,2\kappa\right)\oplus\left(1/2,2\kappa - 1\right),\nn
    {\rm fermions}\qquad &&\left(0,2\kappa - 1/2\right)\oplus\left(1,2\kappa - 1/2\right).
\eea
Here, the first and second numerals in the brackets stand for the spin and $F$ charge of the given component field. The $\Sigma$ charge assignment
of all fields can also be easily established.
\subsubsection{Invariant action}
The general $SU(2|1)$ invariant action can be constructed as
\bea
    S_{\rm kin.}=\int dt\,{\cal L}_{\rm kin.} = \frac{1}{2}\int d\zeta\,f\left(\Phi_i\bar{\Phi}^i\right),\label{kin}
\eea
where the $SU(2|1)$ invariant measure $d\zeta$ is given by
\bea
    d\zeta = dt\,d^2\theta\,d^2\bar{\theta}\left(1+2m\,\bar{\theta}^k\theta_k\right),\quad \delta\left(d\zeta\right)=0\,.\label{measure}
\eea
The Lagrangian in \p{kin} should be a function of the $U(2)$ invariant argument $\Phi_i\bar{\Phi}^i$ due to the presence of the $U(2)$ induced
terms in the transformations \eqref{trPhi}.

The candidate superpotential term can be written as
\bea
    S_{\rm pot}=\tilde{m}\int d\zeta_L\,{\cal F}\left(\phi_i\right)+\mbox{c.c.},
\eea
where the left measure $d\zeta_L = dt_L\,d^2\theta$ is not invariant:
\bea
    \delta\left(d\zeta_L\right)= - \,2m \left(d\zeta_L\right) \bar{\epsilon}^k\theta_k\,.\lb{deltaMeas}
\eea
From \eqref{trPhi}, it follows that there is no way to define the function ${\cal F}\left(\phi_i\right)$ so as to compensate the
measure transformation \p{deltaMeas}. Indeed, the only $SU(2)$ invariant argument $\phi_i\phi^i$ is identically vanishing.
Thus no superpotential action for the $s=1/2$ case can be constructed.\footnote{The
superpotential terms can be written
only for the integer values of $s$ ($s=0,1,2\ldots$) and nonvanishing $\kappa$\,.
 For example, in the $s=1$ case, we deal with the triplet superfield $\phi_{ik}\left(t_L,\theta_k\right)$ defined
 on the left chiral subspace and having the transformation law \p{trPhi} in which the $SU(2)$ part is properly modified
 and the $U(1)$ part appears with the weight $4\alpha\kappa(\bar\epsilon^i\theta_i)$, where $2\alpha\kappa$ is the corresponding $\tilde{F}$ charge.
 The $SU(2)$ invariant argument $\phi_{ik}\phi^{ik}$ undergoes only $U(1)$ transformation:
$\delta \left(\phi_{ik}\phi^{ik}\right) = 8\alpha \kappa m\,\bar{\epsilon}^k\theta_k\,\left(\phi_{ik}\phi^{ik}\right).$
Then, a nonvanishing superpotential for $s=1$ is $ S_{\rm pot}=\int d\zeta_L\,\big(\phi_{ik}\phi^{ik}\big)^{\frac{1}{4 \alpha\kappa}}$.}

\subsubsection{Free model}
The free Lagrangian \eqref{free} with $\gamma=0$ is a part of the following $SU(2|1)$ invariant (modulo a total derivative) free superfield Lagrangian:
\bea
    {\cal L}^{\rm free}_{\rm kin.} = \frac{1}{4}\int d^2\theta\,d^2\bar{\theta}\left(1+2m\,\bar{\theta}^k\theta_k\right)
    \Phi_i \bar{\Phi}^i.\label{freeL}
\eea
To see this, we rewrite \p{freeL} in terms of the component fields:
\bea
    {\cal L}^{\rm free}_{\rm kin.}&=& \bar{\nabla}_t {\bar{z}}^i\,\nabla_t {z}_i+ \frac{i}{2}\left(\bar{\psi}\,\nabla_t\psi
    + \psi\,\bar{\nabla}_t\bar{\psi} \right)
    +\frac{i}{2}\left(\bar{\psi}_{(ik)}\,\nabla_t\psi^{(ik)} + \psi^{(ik)}\,\bar{\nabla}_t\bar{\psi}_{(ik)}\right) +B^i\bar{B}_i\nn
    &&-\,\frac{i}{2}\,m\left(z_i\bar{\nabla}_t {\bar{z}}^i- \bar{z}^i\nabla_t {z}_i\right)
    +\frac{m}{2}\left(\psi\bar{\psi}-3\psi^{(ik)}\bar{\psi}_{(ik)}\right) - \frac{3m^2}{4}\,z_i\bar{z}^i.\label{compL}
\eea
It can be split into a sum of three ${\cal N}=2$ Lagrangians:
\be
{\cal L}^{\rm free}={\cal L}^{\rm free}_{(\rho)} \left.\right|_{\gamma =0}
+{\cal L}^{\rm free}_{(\bf 0, 2, 2)}+{\cal L}^{\rm free}_{(\bf 2, 2, 0)}\,. \lb{s12Decomp}
\ee
Here, ${\cal L}^{\rm free}_{(\rho)} \left.\right|_{\gamma =0}$ is just the $\gamma=0$ version\footnote{The Lagrangians of long multiplets with $\gamma \neq 0$
are contained in the $SU(2|1)$ Lagrangians for the chiral superfields with $s>1/2\,$ (see Sec. 3.2).}  of the long multiplet Lagrangian \p{free}
for the fields $(z, \xi, \pi, B)$ defined in \p{subsets},
with $\sigma = \left(2\kappa - 1/2\right)m$ and $\rho =-\,m$\,. The remaining two Lagrangians are
\bea
{\cal L}_{(\bf 2, 2, 0)}^{\rm free}&=& \bar{D}_t{\bar{\hat{z}}}\,D_t\hat{z}+ \frac{i}{2}\left(\bar{\hat{\xi}}\,D_t\hat{\xi} + \hat{\xi}\,\bar{D}_t\bar{\hat{\xi}}\right)
-\gamma\left[\frac{i}{2}\left(\hat{z}\,\bar{D}_t\bar{\hat{z}}
    -\bar{\hat{z}}\,D_t{\hat{z}}\right) +\hat{\xi}\bar{\hat{\xi}}\right],
\eea
with $\hat{z} = z_2$\,, $\hat{\xi} = -\,\psi^{(11)}$, $\sigma = \left(2\kappa +1/2\right)m$\,, $\gamma = 2m$\,, and
\bea
{\cal L}^{\rm free}_{(\bf 0, 2, 2)}=\frac{i}{2}\left(\bar{\hat{\pi}}\,D_t\hat{\pi} + \hat{\pi}\, \bar{D}_t\bar{\hat{\pi}}\right) + \hat{B}\bar{\hat{B}}\,,
\eea
with $\hat{\pi} = \psi^{(22)}$\,, $\hat{B} = B^2$, $\sigma = \left(2\kappa -3/2\right)m$\,.

\subsubsection{Quantum generators}
Let us consider the quantum system corresponding to the Lagrangian \eqref{compL}.
Quantized brackets are
\bea
    &&\left[z_j,p^i\right]=i\delta^i_j\,,\qquad\left[\bar{z}^i,\bar{p}_j\right]=i\delta^i_j\,,\nn
    &&\left\lbrace\psi ,\bar{\psi}\right\rbrace =1\,,\qquad\left\lbrace\psi^{(ik)}\,,
    \bar{\psi}_{(jl)}\right\rbrace=\frac{1}{2}\left(\delta^i_j\delta^k_l + \delta^i_l\delta^k_j\right).
\eea
The relevant quantum Hamiltonian reads
\bea
    H &=& \left[p^i+i\left(2\kappa -\frac{1}{2}\right)m\bar{z}^i\right]\left[\bar{p}_i-i\left(2\kappa -\frac{1}{2}\right)mz_i\right]
    - \left(2\kappa -\frac{3}{2}\right)\left(2\kappa +\frac{1}{2}\right) m^2z_i\bar{z}^i\nn
    &&-\,\left(2\kappa +\frac{1}{2}\right)m\,\psi\bar{\psi}-\left(2\kappa - \frac{3}{2}\right)m\,\psi^{(ik)}\bar{\psi}_{(ik)}\,.\label{H1}
\eea
Quantum supercharges are given by the expressions
\bea
    &&Q^i=\psi\left(p^i +im\bar{z}^i\right)+\sqrt{2}\,\psi^i_k\left(p^k-im\bar{z}^k\right),\nn
    &&\bar{Q}_j=\bar{\psi}\left(\bar{p}_j-imz_j\right)-\sqrt{2}\,\bar{\psi}^k_j\left(\bar{p}_k+im z_k\right),
\eea
while the bosonic generators are given by
\bea
    &&F=-\,2i\kappa\left(z_k p^k-\bar{z}^k\bar{p}_k\right)-\left(2\kappa -
    \frac{1}{2}\right)\left(\psi\bar{\psi}+\psi^{(ik)}\bar{\psi}_{(ik)}\right),\nn
    &&I^i_j = -\,i\left(z_j p^i - \bar{z}^i\bar{p}_j\right)
    +\frac{i}{2}\,\delta^i_j\left(z_k p^k-\bar{z}^k\bar{p}_k\right)+2\,\psi^{(ik)}\bar{\psi}_{(jk)}-\delta^i_j\,\psi^{(lk)}\bar{\psi}_{(lk)}\,.
\eea
The generators of the ${\cal N}=2$ subalgebra \eqref{algebraSigma} are realized as
\bea
    Q^1&=&\psi\left(p^1 +im\bar{z}^1\right)+\sqrt{2}\,\psi^1_k\left(p^k-im\bar{z}^k\right),\nn
    \bar{Q}_1&=&\bar{\psi}\left(\bar{p}_1-imz_1\right)-\sqrt{2}\,\bar{\psi}^k_1\left(\bar{p}_k+im z_k\right),\nn
    \Sigma &=&-\,i\left(2\kappa - \frac{1}{2}\right)\left(z_1 p^1-\bar{z}^1\bar{p}_1\right)-\left(2\kappa
    - \frac{1}{2}\right)\left(\psi\bar{\psi}+2\,\psi^{(12)}\bar{\psi}_{(12)}\right)\nn
    &&-\,i\left(2\kappa + \frac{1}{2}\right)\left(z_2 p^2-\bar{z}^2\bar{p}_2\right)-\left(2\kappa
    + \frac{1}{2}\right)\psi^{(11)}\bar{\psi}_{(11)}
    -\left(2\kappa - \frac{3}{2}\right)\psi^{(22)}\bar{\psi}_{(22)}.\nn
\eea

Under the condition
\bea
    \left|4\kappa - 1\right| \leq 2\,,\label{Bound1}
\eea
the spectrum of Hamiltonian \eqref{H1} is bounded from below, since it is equivalent to the condition \eqref{Bound} for the long multiplet
with $\sigma = \left(2\kappa -1/2\right)m$\,, $\gamma = 0$\,, $\rho =-\,m$ and for the short multiplet ${\bf(2,2,0)}$ with
$\sigma = \left(2\kappa +1/2\right)m$\,, $\gamma = 2m$\,, $\rho =0$\,. The equation \p{Bound1} is similar to the analogous condition for $s=0$, $\left|4\kappa - 1\right| \leq 1\,$,
deduced in \cite{DSQM}.
%The spectrum of the multiplet ${\bf(0,2,2)}$
%The characteristic parameters of these multiplets are written in
%The Hilbert space of the multiplet ${\bf(0,2,2)}$ is given by only 2 states.

\subsection{Generic $s$}
In the general case of $s\in \mathbb{Z}, \mathbb{Z} + 1/2$, the chiral superfield $\Phi_{(i_1\,\ldots \,i_{2s})}$ belongs
to the $U(2)$ representation $\left(s,2\kappa\right)$ and is defined by the constraints
\footnote{The values of the external $\tilde{F}$ charge can be different for different $s$. We choose them equal for simplicity.}
\bea
    &&\bar{{\cal D}}_j\Phi_{(i_1\,\ldots\, i_{2s})} = 0\,,\qquad  \tilde{F}\Phi_{(i_1\,\ldots\, i_{2s})} = 2\kappa\,\Phi_{(i_1\,\ldots\, i_{2s})}\,,\nn
    &&\tilde{I}^{k}_{l}\Phi_{(i_1\,\ldots\, i_{2s})} = \sum_{n = 1}^{2s}\left[\delta^k_{i_n}\Phi_{(i_1\,\ldots\, i_{n-1}\,l\,i_{n+1}\,\ldots\, i_{2s})}
    - \frac{1}{2}\,\delta^k_l\Phi_{(i_1\,\ldots\, i_{2s})}\right].
\eea
The free action is again given by an integral over the $SU(2|1)$ invariant measure \eqref{measure}:
\bea
    S^{\rm free}_{\rm kin.} \sim \int d\zeta\,\Phi_{(i_1\,\ldots\, i_{2s})}\bar{\Phi}^{(i_1\,\ldots\, i_{2s})}.
\eea
The general action involves an arbitrary $U(2)$ invariant Lagrangian function of $\Phi_{(i_1\,\ldots \,i_{2s})}$.

The superfield $\Phi_{(i_1\,\ldots \,i_{2s})}$ amounts to the component field expansion with the following $U(2)$ assignments
\bea
    {\rm bosons}\qquad &&\left(s,2\kappa\right)\oplus\left(s,2\kappa - 1\right),\nn
    {\rm fermions}\qquad &&\left(s-1/2,2\kappa - 1/2\right)\oplus\left(s+1/2,2\kappa - 1/2\right).
\eea
On all these fields, the $U(1)$ generator $\Sigma$  takes the values $\sigma$ in the range between $\left(2\kappa-1-s\right)m$
and $\left(2\kappa+s\right)m$\,.
%\footnote{Here, $0\leqslant2\kappa\leqslant1/2$ and $m>0$\,.}
For the minimal and maximal values of $\sigma$ we observe the 2-fold degeneracy of fields (one complex boson and one complex fermion).
%with respect
%to the generator $\Sigma$\,.
These two sets are described by the irreducible chiral ${\cal N}=2$ superfields $({\bf 2, 2, 0})$ and $({\bf 0, 2, 2})\,$.
All other complex fields have the following values of $\sigma$:
\bea
    \sigma = \left(2\kappa - s - 1 +n\right)m\,,\qquad n=1,2,\ldots 2s\,.
\eea
The relevant fields exhibit the four-fold degeneracy (two complex bosons and two complex fermions).
Every such a four-fold set forms a long multiplet.

Thus, the chiral $SU(2|1)$ multiplet with the external quantum numbers $\left(s,2\kappa\right)$ decomposes into the following
sum of ${\cal N}=2$ multiplets:
\bea
2s \;{\rm long \;multiplets} \oplus  {\rm one \; short} \;({\bf 2, 2, 0})\; {\rm multiplet} \oplus  {\rm one\; short} \; 
({\bf 0, 2, 2}) \; {\rm multiplet}.\lb{DecoM}
\eea

For the fixed  long multiplet from this decomposition, the mass-dimension parameters $\rho$ and $\sigma$ are proportional to $m$ and are
specified by the numbers $s$, $2\kappa$, and $n$. The external charges of ${\cal N}=2$ superfields corresponding to the long multiplets
are $\sigma_{(s,n)} = \left(2\kappa - s- 1 +n\right)m\,$, with $n=1,2,\ldots 2s\,.$ For the superfields corresponding to the short multiplets
in \p{DecoM}, the values of $\sigma$ are $\left(2\kappa + s\right)m$ and $\left(2\kappa -s -1\right)m\,$. All these ${\cal N}=2$ superfields
are defined either by the constraints \p{NewLp}, with $\rho_{(s,n)} = -\sqrt{n\left(2s + 1 -n\right)}\,m, \;n=1,\ldots, 2s,$ 
or by the standard chirality constraints,
with $\bar{\cal D}=-\,\partial/\partial \bar{\theta}+i\theta\partial_t - \theta \sigma$. For $s=1/2$, we have $\rho = -\,m$\,, 
in agreement with Sec. 3.1.
For the generic $s$, one also finds that $\gamma_{(s,n)} = \left(2n- 2s- 1\right)m$\,,\; $n=1,\ldots, 2s+1\,,$ leading to
the generalization of the condition \eqref{Bound1}:
\bea
    \left|4\kappa - 1\right| \leq 2s + 1.
\eea
It amounts to \p{Bound} for each ${\cal N}=2$ multiplet. Note that the $s=0$ model \cite{DSQM} involves the cases $\kappa=0$
and $\kappa=1/2$ with the degenerate vacuum states, for which $\left|4\kappa - 1\right| = 1$. This matches with
the extra vacuum degeneracy in the case of equality in \p{Bound}.

The general formulas for the parameters $\rho_{(s,n)}$ and $\gamma_{(s,n)}$ were found from the requirement that the free Lagrangians of long multiplets
have the generic form \p{free}.
To obey this requirement, one should, in particular, properly rescale
those linear combinations of the original component fields of $\Phi_{(i_1\,\ldots \,i_{2s})}$ which are going to become the quotient fields
$(\pi, B)$ (no such a rescaling is needed for $s=1/2$).

\subsection{Dimensional reduction}
The authors of \cite{Casimir} considered a dimensional reduction of the free $d=4$ model of the chiral multiplet
on $S^3\times \mathbb{R}$\, \cite{FS}. After reduction to $d=1$\,, it became the free model of $SU(2|1)$ mechanics
with an infinite set of fields\footnote{In \cite{Casimir}, all fields also have an additional index of
the external group $SU(2)_{r}$ of the Pauli-G\"ursey type. Here, we do not consider such a group.}
 corresponding to a sum of long multiplets and the chiral multiplets $({\bf 2, 2, 0})$ and $({\bf 0, 2, 2})$.
Comparing this set with the ${\cal N}=2$ decomposition of $SU(2|1)$ chiral superfields, one concludes that the relevant SQM system
involves a sum of $SU(2|1)$ chiral supermultiplets with all integer and half-integer spins ${s} = 0, 1/2, 1, \ldots\,. $
They are described by the chiral $SU(2|1)$ superfields carrying the symmetrized multi-indices $(i_1\cdots i_{2s})$, which corresponds
to the harmonic expansion on $S^3$.

It is worth noting that our intrinsically one-dimensional superfield approach  allowed us not only to straightforwardly
reproduce the free actions for such superfields but also to construct their most general $SU(2|1)$ invariant self-interactions,
including the Wess-Zumino and superpotential terms. While in the explicit example of Sec. 3.1 we limited our study to the simplest case of the external
spin ${s}=1/2$, there is no obstacle against including higher ${s}$ superfields and constructing general interactions among them.
It would be interesting to use these techniques to find possible corrections to the vacuum Casimir energies computed
in \cite{Casimir} at the level of free $SU(2|1)$ invariant actions.

\setcounter{equation}{0}
\section{Outlook}
We showed that the long multiplets naturally appear in $SU(2|1)$ mechanics of chiral multiplets \cite{DSQM},
when the chiral superfield $\Phi_A$ carries some external index $A$ with respect to the stability subgroup $SU(2)$
of the chiral ${\cal N}=4$\,, $d=1$ superspace constructed on the basis of the centrally extended supergroup $SU(2|1)$.
We used the superfield techniques to construct the most general actions of the long multiplets,
in the framework of both ${\cal N}=2$ supersymmetry and the $SU(2|1)$ extension of the latter. As a byproduct, we
revealed the existence of another type of long ${\cal N}=2, d=1$ multiplets, with the opposite overall Grassmann
parity as compared to that considered in \cite{Casimir}.

In \cite{SKO}, we introduced another type of $SU(2|1)$ chiral superfields, with the generalized chiral constraints involving
a new dimensionless parameter $\lambda$\,. Unfortunately, such constraints are self-consistent  only for the chiral superfields carrying no external $SU(2)$ spin.
Yet the external indices of $SU(2)\subset SU(2|1)$ can be ascribed to some other $SU(2|1)$ superfields, with the field contents $({\bf 1, 4, 3})$, $({\bf 4,4,0})$ and
$({\bf 3, 4, 1})$ \cite{DSQM, DHSS}. It would be of obvious interest  to analyze the relevant decompositions with respect
to ${\cal N}=2$ supersymmetry.

In \cite{Casimir}, the one-dimensional quantum-mechanical problem for a single long multiplet was solved.
It would be interesting to solve the $SU(2|1)$ extension of this problem associated with the free Lagrangian \eqref{freeL}
and to analyze the role of the long multiplet contributions from the standpoint of the $SU(2|1)$ representation theory.
In contrast to \eqref{compL}, the general action \eqref{kin} involves interaction terms of ${\cal N}=2$ multiplets.
Hence, it cannot be considered as a sum of free actions of ${\cal N}=2$ multiplets. As was already mentioned,
it would be  tempting to find out possible physical effects of these interactions, e.g., in the context of consideration
in ref. \cite{Casimir}.

In conclusion, we briefly discuss how ${\cal N}=2, d=1$ long multiplets can be generalized to the case of the standard flat ${\cal N}=4$, $d=1$ supersymmetry.

One of the options is to consider a system of complex  ${\cal N}=4$ superfields $\Lambda$ and $Z_i, i =1,2,$ subjected
to the constraints
\bea
    \bar{D}_{i}\Lambda = -\,\sqrt{2}\,r\,Z_i\,,\qquad \bar{D}^{i}\bar{D}_{i}\Lambda = 0\;\Rightarrow \;  \bar{D}_{k}Z_i = 0\,. \lb{N4A}
    %\qquad \lbrace D^i, \bar{D}_k\rbrace = 2i\delta^i_k\partial_t\,,
\eea
The superfield  $\Lambda$ can be fermionic or bosonic, while $Z_i$ should have the Grassmann parity opposite to $\Lambda$.
The constraints \p{N4A} are solved through the set of chiral ${\cal N}=4$\,, $d=1$ superfields $\Lambda_{0}$ and $Z_i$:
\be
    \Lambda = \Lambda_{0} + \sqrt{2}\,r\,\bar{\theta}^i Z_i\,,\qquad\bar{D}_{k}\Lambda_{0} = 0\,,\qquad \bar{D}_{k} Z_i = 0\,.
\ee
These superfields transform as\footnote{We use the ``passive'' form of the transformations.}
\be
    \delta\Lambda = 0\,, \; \delta Z_i = 0 \quad\Rightarrow\quad\delta\Lambda_{0} = -\,\sqrt{2}\,r\,\bar{\epsilon}^i Z_i\,,
\ee
implying that the component  transformations contain the parameter $r$ and so correspond to an indecomposable multiplet.
In the limiting case $r=0$\,, the superfield $\Lambda$ becomes just a standard chiral ${\cal N}=4$ superfield.

Another indecomposable ${\cal N}=4$ supermultiplet parametrized by a real dimensionless parameter $\alpha$ was suggested in \cite{Katona} in the component approach.
As a matter of fact, the same system in the superfield approach can be defined by the constraints:
\bea
    \bar{D}_k V = \frac{i\alpha}{2}\,D_k W\,,\qquad \bar{D}_k W = 0\,,\label{katona}
\eea
where $V$ and $W$ are bosonic superfields. Equations \p{katona}  are solved by
\bea
    V = V_0 - \frac{i\alpha}{2}\,\bar{\theta}^k D_k W +\frac{\alpha}{2}\,\big(\bar\theta\,\big)^2\dot{W}\,,\qquad \bar{D}_k V_0 = 0\,,
\eea
implying  the following transformation properties  of the superfields involved:
\bea
    \delta V_0 =  \frac{i\alpha}{2}\, \bar{\epsilon}^k D_k W -\alpha\left(\bar{\epsilon}^k\bar{\theta}_k\right)\dot{W}\,,\qquad
\delta W = \delta V = 0\,.
\eea
In the limit $\alpha=0$\,, the constraints \eqref{katona} are reduced to those defining two ordinary $({\bf 2, 4, 2})$
chiral multiplets. First, the constraints \eqref{katona} can be generalized by adding some $SU(2)$ breaking term $\sim c_{(k}^{\;\;\;\;l)}D_l W$ to
its right-hand side. One can choose the $SU(2)$ frame so that  $c_{(k}^{\;\;\;\;l)} = \beta (\tau_3)_{k}^{\;l}$. Presumably, this will give rise
to the two-parameter deformed multiplet also considered in  \cite{Katona}.\footnote{A different type of indecomposable ${\cal N}=4, d=1$ multiplets
is provided by the classification scheme of ref. \cite{GonTo}.}

Both sets of the constraints, \p{N4A} and \p{katona}, can be generalized to the $SU(2|1)$ case. Such a generalization of \p{katona}
is possible only for the second type of the  $SU(2|1)$ chirality \cite{SKO}, when the spinor derivatives are inert under the induced
$U(1)$ transformations.

It would be also tempting to reveal $d>1$ analogs of the long multiplets considered in the present paper.

\section*{Acknowledgements}
We acknowledge support from the RFBR grant No. 15-02-06670 and a grant of the Heisenberg-Landau program. We are grateful to Sergey Fedoruk
and Francesco Toppan for useful discussions.

\end{document}